\documentclass[a4paper]{jpconf}
\usepackage{graphicx}
\begin{document}
\def\bb    #1{\hbox{\boldmath${#1}$}}

\title{Analytical Expressions for the Hard-Scattering Production of
  Massive Partons}

\author{Cheuk-Yin Wong} 

\address{Physics Division, Oak Ridge National Laboratory,
Oak Ridge, Tennessee 37831, USA}


\begin{abstract}
 We obtain explicit expressions for the two-particle differential
 cross section and the two-particle angular correlation function 
  in the
 hard-scattering production of massive partons in order to exhibit the
 ``ridge" structure on the away side in the hard-scattering process.
 The single-particle production cross section is also obtained and compared with the ALICE
 experimental data for charm production in pp collisions at 7 TeV at
 LHC.

\end{abstract}

\vspace*{-0.8cm}
\section{Introduction}

Knowledge of massive quark production processes in $pp$ collisions
provides useful insight to guide our intuition in heavy quark
production in nucleus-nucleus collisions.  Analytical expressions for
these processes summarize important features and essential
dependencies so as to facilitate the uncovering of dynamical effects
wherever they may occur.  Similar analyses in massless quark
production have led to new insights in the dominance of the
hard-scattering process over a large $p_T$ domain and have paved the
way for locating the boundary between the hard-scattering process and
the flux-tube fragmentation process in high-energy $pp$ collisions \cite{Won12,Won15c}.

Accordingly, we would like to obtain $E_c E_\kappa d\sigma (AB$$ \to$$
c\kappa X) /d\bb c \,d \bb \kappa$ for the two-particle differential
cross section in the production of massive partons $c$ and $\kappa$.
From such a general result, we integrate out the transverse momenta
and obtain the two-particle angular correlation function
$d\sigma/d\Delta \phi d\Delta y$ where $\Delta
\phi$=$\phi_{\kappa}$$-$$\phi_c $ and $\Delta y$=$y_{\kappa}$$-$$ y_c
$, exhibiting analytically the ``ridge" structure on the away side at
$\Delta \phi$$\sim $$\pm \pi$ in the hard-scattering process.  We
subsequently examine $d\sigma(AB $$\to$$ cX) /dy_c c_T dc_T $ for the
single-particle spectrum and compare with ALICE experimental data for
charm production in $pp$ collisions at 7 TeV at LHC \cite{ALICE12}.

\section{Hard Scattering Integral for $E_c E_\kappa d\sigma(AB\to c\kappa X) /d\bb c \, d\bb \kappa$ }

In the parton model, the hard-scattering cross section for $AB\to c
\kappa X$ is given by
\cite{Owe87}
\begin{eqnarray}
d\sigma ( AB \to c \kappa X) 
=\sum_{ab} \int K_{ab} dx_a d{\bb a}_T dx_b d{\bb b}_T
G_{a/A}(x_a,{\bb a}_T) G_{b/B} (x_b,{\bb b}_T) d\sigma( ab \to c \kappa) ,
\end{eqnarray}
where $(x_a, \bb a_T) $ and $(x_b, \bb b_T) $ represent the momenta
and $G_{a/A}$ and $G_{b/B}$ the structure functions of the incident
partons $a$ and $b$ respectively, and $K_{ab}$ is the correction
factor which can be obtained perturbatively \cite{Sjo15} or it can
also be approximated nonperturbatively \cite{Cha95}.  The quantity
$d\sigma( ab \to c \kappa)$ is the cross section element for the
process
$ab \to c\kappa$,
\begin{eqnarray}
d\sigma (ab\to c\kappa)= \frac{ 1 }{4[(a \cdot b)^2 - m_a^2 m_b^2 ] ^{1/2}} |{ T}_{fi}|^2 
 \frac{d^3c}{(2\pi)^3 2 E_ c} \frac{ d^3\bb \kappa}{(2\pi)^3 2 E_ {\kappa}}
(2 \pi)^4 \delta^4 ( a+b- c-\kappa).
\end{eqnarray}
Here, we normalize the Dirac fields by $\bar u u = 2m$.  The quantity
$|T_{fi}|^2$ is related to $d\sigma/dt$ by
\begin{eqnarray}
|T_{fi}|^2 = 16 \pi [\hat s-(m_a+m_b)^2][\hat s-(m_a-m_b)^2]  \frac{d\sigma(ab\to c\kappa) }{dt}.
\label{8}
\end{eqnarray}
We consider the simplified case with $m_a=m_b=0$ and treat $a_T, b_T$
as small perturbations.  The cross section element is then
\begin{eqnarray}
d\sigma(ab\to c\kappa)
= \frac{ s_{ab}}{2\pi}\! \frac{d\sigma(ab\to c\kappa) }{dt}
 \frac{d^3c}{ E_ c} \frac{d^3\kappa}{ E_ \kappa}
\delta^4 ( a+b - c - \kappa),
\end{eqnarray}
where $\hat s=s_{ab}=(a+b)^2$ that is different from
$s=s_{AB}=(A+B)^2$.  We get
 \begin{eqnarray}
\frac{E_c E_\kappa d\sigma (\! AB\!\! \to\!\!c \kappa X\!)}{d^3 c ~ d^3 \kappa}  
\!=\!\sum_{ab}\!\!\! \int  \!\!K_{ab} dx_a d{\bb a}_T dx_b d{\bb b}_T
G_{a/A}\!(x_a,\!{\bb a}_T\!) G_{b/B}\! (x_b,\!{\bb b}_T\!) 
\frac{ \hat  s}{2\pi}\!\!\frac{d\sigma \!(ab\!\to\! c\kappa) }{dt}
\delta^4\! ( a\!+\!b\! -\!c \!-\! \kappa).
\label{eq5}
\end{eqnarray}
We consider a fractorizable structure function with a Gaussian
intrinsic transverse momentum distribution,
\begin{eqnarray}
G_{a/A}(x_a,{\bb a}_T)=G_{a/A}(x_a) \frac{1}{2\pi \sigma^2} e^{-{\bb a}_T^2/2\sigma^2} .
\end{eqnarray}
Upon integrating over the transverse momenta $\bb a_T$ and $\bb b_T$,
we obtain
\begin{eqnarray}
\frac{d\sigma ( AB \to c \kappa X)}{dy_c c_T dc_T d \phi_c\, dy_\kappa \kappa_T d\kappa_T d \phi_\kappa  }  
&=&\sum_{ab} \int K_{ab} dx_a  dx_b 
G_{a/A}(x_a)G_{b/B} (x_b) 
 \frac{ e^{-\frac{(\bb c_T + \bb \kappa_T)^2}{4\sigma^2}}} {2(2\pi \sigma^2)}
\frac{ \hat  s}{2\pi}\frac{d\sigma(ab\to c\kappa) }{dt}
\nonumber\\
& &\times 
\delta ( a_0+b_0 - (c_0+\kappa_0))\delta ( a_z+b_z - (c_z+\kappa_z)).
\label{eq8} 
\end{eqnarray}
To carry out the integration over $x_a$ and $x_b$, we write out the
momenta in the infinite momentum frame,
\begin{eqnarray}
a&=&(x_a \frac{\sqrt{s}}{2} + \frac{a^2+a_T^2}{2x_a \sqrt{s}}, ~{\bb a}_T, ~~x_a \frac{\sqrt{s}}{2} - \frac{a^2+a_T^2}{2x_a \sqrt{s}}),
\\
b&=&(x_b \frac{\sqrt{s}}{2} + \frac{b^2+b_T^2}{2x_b \sqrt{s}}, ~\, {\bb b}_T,-x_b \frac{\sqrt{s}}{2} + \frac{b^2+ b_T^2}{2x_b \sqrt{s}}),
\\
c&=&(x_c \frac{\sqrt{s}}{2} + \frac{c^2+c_{T}^2}{2x_c \sqrt{s}}, ~{\bb c}_T, ~~x_c \frac{\sqrt{s}}{2} - \frac{c^2+c_T^2}{2x_c \sqrt{s}}) ,
\\
\kappa&=&(x_\kappa \frac{\sqrt{s}}{2} + \frac{\kappa^2+\kappa_T^2}{2x_\kappa \sqrt{s}},\,{\bb \kappa}_T, -x_\kappa \frac{\sqrt{s}}{2} + \frac{\kappa^2+\kappa_T^2}{2x_\kappa \sqrt{s}}) ,
\end{eqnarray}
where  $x_c$ and $x_\kappa$ can be  represented by $y_c$ and $ y_\kappa$ 
\begin{eqnarray}
x_c =\frac{m_{cT} e^{y_c} }
{\sqrt{s}},   ~~~
x_\kappa =  \frac{m_{\kappa T} e^{y_\kappa} }{\sqrt{s}}.
\end{eqnarray}
The two delta functions in Eq.\ (\ref{eq8}) can be integrated to yield
\begin{eqnarray}
\frac{d\sigma ( AB \to c d X)}{dy_c c_T dc_T d \phi_c\, dy_\kappa \kappa_T d\kappa_T
 d \phi_\kappa  }  =\sum_{ab} 
K_{ab} x_{a}G_{a/A}(x_{a}) x_{b}G_{b/B} (x_{b}) 
 \frac{ e^{-\frac{(\bb c_T + \bb d_T)^2}{4\sigma^2}}} {2\pi (4\pi \sigma^2)}
\frac{d\sigma(ab\to  c\kappa) }{dt},
\label{eq13}
\\
{\rm where} ~~~~~~~
\hspace{0.0cm}x_a= x_c + \frac{\kappa^2+\kappa_T^2}{x_\kappa s } - \frac{b^2+ b_T^2}{x_b s}
= \frac{m_{cT} e^{y_c} }{\sqrt{s}} + \frac{m_{\kappa T} e^{-y_\kappa } }{\sqrt{s}} - \frac{b^2+b_T^2}{x_b s},
\nonumber\\
x_b
=x_\kappa  +\frac{c^2+c_{T}^2}{x_c {s}}-\frac{a^2+a_T^2}{x_a s}
=\frac{m_{\kappa T}e^{y_\kappa}}{\sqrt{s}}  +\frac{m_{cT}e^{ -y_c} }{\sqrt{s}}
-\frac{a^2+T^2}{x_a s}.
\end{eqnarray}
The above explicit formula gives the cross section for the production
of $c$ and $\kappa$, when the elementary cross section ${d\sigma(ab\to
  c\kappa) }/{dt}$ is given explicitly in terms of its depending
variables.

\section{ The angular correlation 
$d\sigma(AB \to c\kappa X)/d\Delta \phi \,d\Delta y$}

We can represent $\bb c$ and $\bb \kappa$ by $(y_c,\phi_c)$ and $(y_c
+\Delta y, \phi_c+\Delta \phi)$, respectively.  After averaging over
$y_c$ and $\phi_c$, and integrating over $c_T, \kappa_T$, the
correlation function (\ref{eq13}) from the process $ab$$\to$$c\kappa$
is
\begin{eqnarray}
\hspace*{-5.5cm}
\frac{d\sigma ( AB \to c \kappa X)}{ d\Delta \phi \, d\Delta y\,   }  =
K_{ab} x_a G_{a/A}(x_{a}) x_b G_{b/B} (x_{b})  \delta_\sigma( \Delta \phi ),~~~~~~
\label{eq14}
\end{eqnarray}
\begin{eqnarray}
{\rm where~~} \delta_\sigma( \Delta \phi )
= \frac{1} { (4\pi \sigma^2)} \int_0^\infty  \!\!c_T dc_T   \, \int _0^\infty\!\!  \kappa_T d\kappa_T   \exp\{-\frac{c_T^2 + 2 c_T \kappa_T \cos \Delta \phi + \kappa^2}{4\sigma^2}\}
\frac{d\sigma(ab \to  c{\kappa}) }{dt}.
\label{eq16}
\end{eqnarray}
The above analytical expression assumes a simple form for $a$=$b$ and
$c$=$\kappa$.  The structure function can be represented in the form $
x_a G_{a/A}(x_{a}) $$\propto $$(1-x_a)^{g_a} $ for which the
two-particle angular correlation function becomes
\begin{eqnarray}
\frac{d\sigma ( AB  \!\to  \!c \kappa X)}{ d\Delta \phi \, d\Delta y\,   } \! \sim \!
 A \! \left [  \!
1 \!- \!\frac{2m_{cT}}{\sqrt{s}} [\cosh y_c  \!+  \! \cosh ( y_c \!+ \!\Delta y)]
 \!+ \!2(\frac{m_{cT}}{\sqrt{s}})^2 [ 1 \!+ \! \cosh ( 2 y_c \!+ \!\Delta y)]
\right ] ^{g_a}  \! \! \!\delta_\sigma( \Delta \phi).
\end{eqnarray}

\vspace*{-0.3cm}
\begin{figure}[h]
\hspace*{0.5cm}
\includegraphics[scale=0.35]{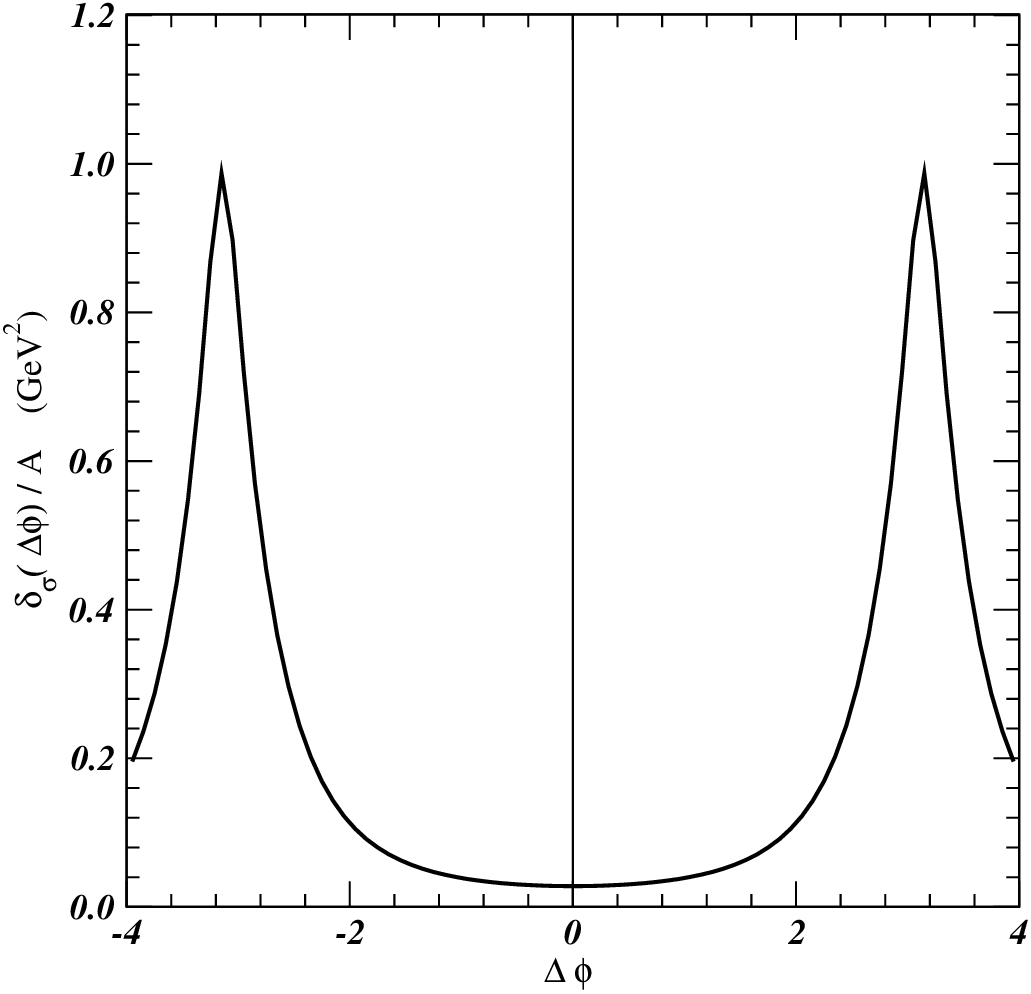}
\end{figure}

\vspace*{-0.3cm}\hspace*{-0.3cm}
{\bf Fig. 1} The  correlation function $\delta_\sigma(\Delta \phi)/A$

\vspace*{-6.3cm}
\hangafter=0
\hangindent=3.3in
\noindent 
If we consider $d\sigma(ab$$ \to$$ c\kappa)/dt$ to be approximately of the form
\begin{eqnarray}
\frac{d\sigma(ab \to c\kappa)}{dt}
=
\frac{A}{ (1+ c_T^2/m_c^2 )^{n/2}},  
\end{eqnarray}
where $n$=4 from pQCD, then the integration over $c_T$ and $\kappa_T$
in Eq.\ (\ref{eq16}) gives $\delta_\sigma(\Delta \phi)/A$ as shown in
Fig. 1.  The correlation function has maxima at $\Delta\phi \sim\pm
\pi$ and a minimum at $\Delta \phi$=0.  It  is relatively flat in $\Delta y$ because $m_{cT}/\sqrt{s} \ll
1$.  This gives the distribution in the form of a ridge structure on
the away side at $\Delta\phi\sim \pm \pi$.

\vspace*{0.6cm}
\section{Production of massive quarks by gluons}

We consider the process $gg \to c\bar c$, where analytical expressions
$d\sigma/dt$ have been obtained earlier by Cambridge \cite{Com79} and
Gl\"uck, Owen, and Reya \cite{Glu78}.  In the notation of
\cite{Glu78}, the cross section is
\begin{eqnarray}
\frac{d\sigma(gg\to c \bar c)}{dt} = 
\frac{\pi \alpha_s^2}{64 \hat s^2} \biggl  [
12 M_{ss}+\frac{16}{3}M_{tt}+\frac{16}{3}M_{uu}+6M_{st}+6M_{su}-\frac{2}{3} M_{tu} \biggr ] .
\end{eqnarray}
Upon writing out the above quantity as a function of $c_T, \bar c_T,
\bar y=(y_c + y_{\bar c})/2$ and $\Delta y $= $y_{\bar c}- y_c$, we
get the heavy-quark pair production cross section
\begin{eqnarray}
\hspace*{-2.1cm}
\frac{d\sigma ( AB \to c \bar c X)}{dy_c c_T dc_T d \phi_c\, dy_{\bar c} \bar c_T d\bar c_T
 d \phi_\kappa  }  \sim A K_{ab} (1-x_a)^{g_a} (1-x_b)^{g_b}
 \frac{ e^{-\frac{(\bb c_T + \bb {\bar c}_T)^2}{4\sigma^2}}} {2\pi (4\pi \sigma^2)}
\frac{d\sigma(gg \to c\bar c)}{dt},
\end{eqnarray}
\begin{eqnarray}
{\rm where~} \frac{d\sigma(gg \!\to \!c\bar c)}{dt}\! =\! \frac{\pi \alpha_s^2}{4^5 m_{cT}^4  \cosh^4 \bar y }
\biggl  \{
 \left [\frac{12}{ \cosh^2 \bar y}\!+\!\frac{64}{3} \cosh 2\bar y\!-\!24   \right ]
\!+\!( \frac{m_c}{m_{cT}})^2\left [\frac{64}{3}\!+ \!\frac{24\sinh^2 \bar y}{\cosh^2 \bar y}-\frac{8}{3}\right  ]\nonumber\\
\hspace*{3.6cm}+\frac{m_c^4}{ m_{cT}^4}  \left  [ -\frac{64}{3}\frac{\cosh 2\bar y}{\cosh^2 \bar y}+\frac{8}{3}\frac{1}{   \cosh^2 \bar y }\right  ]\biggr  \},\nonumber
\end{eqnarray}
$x_a=2 m_{cT} \cosh \bar y e^{\Delta y/2}$, and $x_b=2 m_{cT} \cosh
\bar y e^{-\Delta y/2}$.  This shows the back-to-back correlation of
$\bb c_T $ and $\bb {\bar c}_T$ and the relatively flat distribution
as a function of $\Delta y$.

\section{ Single-particle charm production}

We need to integrate over $\bb \kappa$ in Eq.\ (\ref{eq5}) to get the
single-particle distribution of $\bb c$.  Neglecting intrinsic $p_T$
and integrating over $\bb \kappa$, we get
 \begin{eqnarray}
\frac{E_c d\sigma ( AB \to c X)}{d^3 c }  =\sum_{ab} \int dx_a  dx_b 
G_{a/A}(x_a) G_{b/B} (x_b) 
\frac{ \hat  s}{\pi}\frac{d\sigma(ab\to cd) }{dt}
\delta(\hat s +\hat t +\hat u-m_c^2 - m_\kappa^2).
\end{eqnarray}
The integral over $x_a$ and $x_b$ can be evaluated by the saddle point
method \cite{Won12}, and we get for $\bar y=y_c\approx 0$,
\begin{eqnarray}
&&\hspace*{-0.8cm}\frac{E_c d\sigma ( AB \to c  X)}{d^3c }  \propto  K_{ab} (1-x_a)^{g_a+1/2} (1-x_b)^{g_b+1/2} \frac{1}{\sqrt{x_c}}
\frac{d\sigma(gg \to c\bar c)}{dt}
\label{eq22}
\\
&&\hspace*{-0.8cm}\! =\! \frac{AK_{ab}\alpha_s^2}{m_{cT}^{4.5}  \!\cosh^4 \!\bar y }
\biggl  \{\!
 \left [\! \frac{12}{ \cosh^2\! \bar y}\!+\!\frac{64}{3} \cosh 2\bar y\!-\!24   \right ]
\!\!+\!\! \frac{m_c^2}{m_{cT}^2}\!\left [\!\frac{64}{3}\!+ \!\frac{24\sinh^2\! \bar y}{\cosh^2 \bar y}\!-\!\frac{8}{3}\!\right  ]\!\!\!
+\!\frac{m_c^4}{ m_{cT}^4}\!  \left  [\! -\frac{64}{3}\frac{\cosh\! 2\bar y}{\cosh^2 \!\bar y}\!+\!\frac{8}{3}\frac{1}{   \cosh^2 \!\bar y }\right  ]\!\biggr  \},\nonumber
\end{eqnarray}

\vspace*{-0.4cm}
\begin{figure}[h]
\includegraphics[scale=0.34]{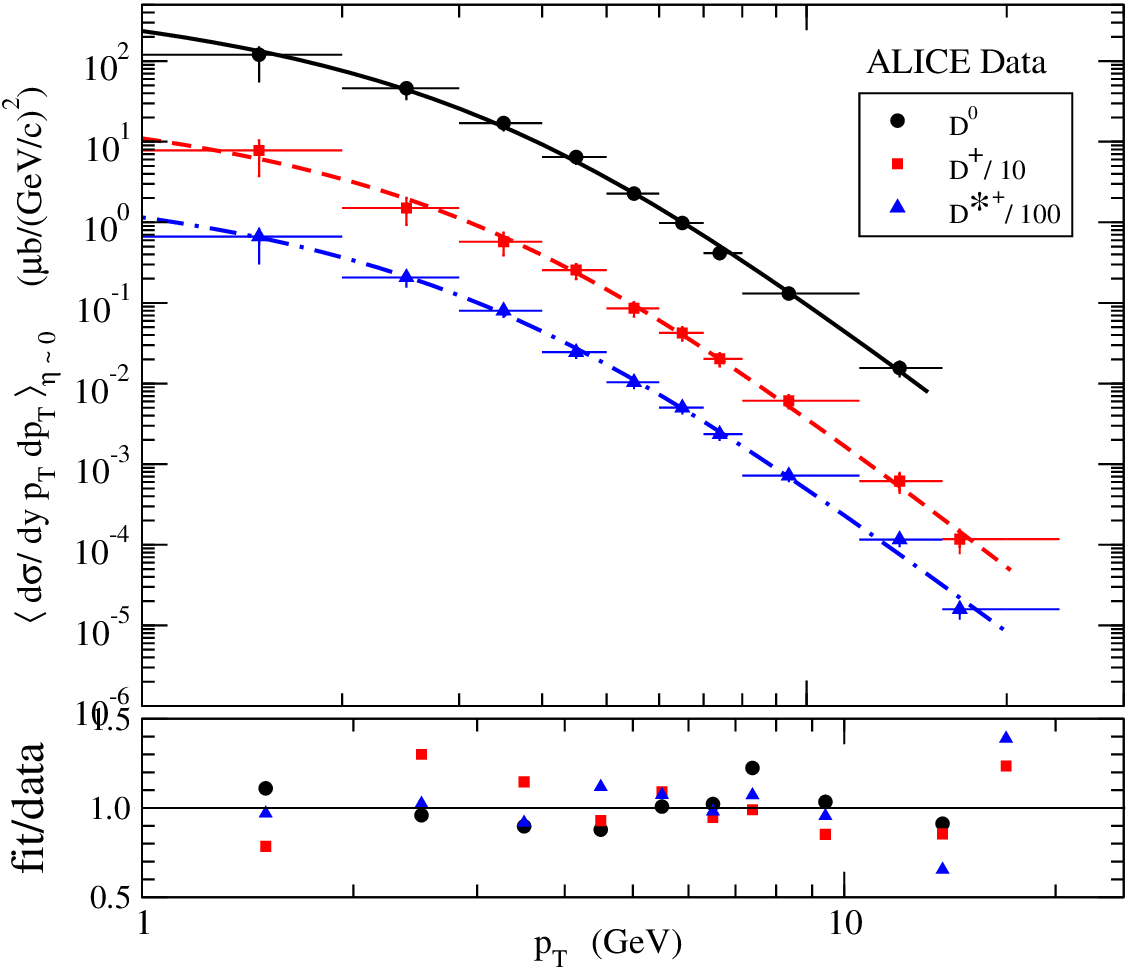}
\end{figure}
\vspace{-0.3cm}
\hspace*{-0.3cm}{\bf Fig. 2} . ALICE  data for charm
 production.

\hangafter=0
\hangindent=3in
\vspace*{-6.1cm}
\noindent 
which contains $1/m_{cT}^4$, $m_c^2/m_{cT}^6$, and $m_c^4/m_{cT}^8$.
We examine the ALICE charm production cross section data in $pp$
collisions at 7 TeV \cite{ALICE12} shown in Fig. 2.  If we parametrize
the data at mid-rapidity as $d\sigma/dy_c c_T dc_T \sim
a/(1+p_T^2/m_0^2)^{n/2}$, we find that the ALICE data can be fitted by
a set of parameters given by
\vspace*{-0.1cm} 
\begin{table}[h]
\hspace*{8.0cm}
\begin{tabular}{|c|c|c|c|}
\cline{1-4}
  \multicolumn{4}{|c|}   {$d\sigma/dy p_Tdp_T|_{\eta\sim 0} =a/(1+p_T^2/m_0^2)^{n/2} $   } 
\\
\cline{1-4}
     Data    &  $a$ [$\mu b$/(GeV/c)$^{-2}$] &  $n$  & $m_0$ (GeV) 
  \\
\cline{1-4}
$D^0$  & 1600 &  5.8 & 3.5
\\
$D^+$ & 780 &  5.9 & 3.5
\\
$D^{*+}$ & 808 &  5.7 & 3.5
\\ \hline
\end{tabular}
\end{table}

\vspace*{0.0cm}
The extracted values of $n$ and $m_0$ are greater than those from the
expected lowest-order results of $n=4.5$ and $m_c\sim1.5$ GeV in
Eq.\ (\ref{eq22}).  This may arise from the final-state interactions
correction factor $K_{ab}$ \cite{Cha95} for production of the
$c$-$\bar c$ pair in the color singlet state.  The attractive
color-singlet interaction between $c$ and $\bar c$ enhances the
production of the pair at lower $p_T$ and increases the value of the
effective mass of the produced charm meson.

\vspace*{-0.2cm}
\section {Conclusion}
We present analytical expressions for the hard-scattering production
of massive quarks in order to guide our intuition, point out essential
dependencies, and summarize important features.  They will facilitate
future comparisons with experimental data and pave the way for a
better understanding of particle production processes.

The research was supported in part by the Division of
Nuclear Physics, U.S. Department of Energy under Contract
DE-AC05-00OR22725.


\end{document}